\documentclass[twocolumn]{aastex631}

\usepackage{amsmath}
\usepackage{booktabs}
\usepackage{graphicx}
\usepackage{url}

\usepackage{siunitx}

\makeatletter
\newcount\LS@mappage \LS@mappage=0
\newcommand{\recordmappage}{%
  \write\@auxout{\string\gdef\string\LS@mappage@val{\thepage}}%
}
\AtBeginDocument{%
  \@ifundefined{LS@mappage@val}{}{%
    \LS@mappage=\LS@mappage@val\relax
  }%
}
\AddToHook{shipout/before}{%
  \ifnum\LS@mappage>0
    \ifnum\value{page}=\LS@mappage
      \pdfpageattr{/Rotate 90}%
    \else
      \pdfpageattr{}%
    \fi
  \fi
}
\def\ps@apjheads{\let\@mkboth\markboth
  \def\@evenfoot{}%
  \def\@evenhead{\lower9mm\hbox to\textwidth{%
    \rm\thepage\hfill\textsc{\@rectohead}\hfill}}%
  \def\@oddfoot{}%
  \def\@oddhead{%
    \ifnum\LS@mappage>0 %
      \ifnum\value{page}=\LS@mappage\hfil\else
        \lower9mm\hbox to\textwidth{%
          \hfil\rm\textsc{\@versohead}\hfil\rm\thepage}%
      \fi
    \else
      \lower9mm\hbox to\textwidth{%
        \hfil\rm\textsc{\@versohead}\hfil\rm\thepage}%
    \fi
  }%
  \def\@evenhead{%
    \ifnum\LS@mappage>0 %
      \ifnum\value{page}=\LS@mappage\hfil\else
        \lower9mm\hbox to\textwidth{%
          \rm\thepage\hfill\textsc{\@rectohead}\hfill}%
      \fi
    \else
      \lower9mm\hbox to\textwidth{%
        \rm\thepage\hfill\textsc{\@rectohead}\hfill}%
    \fi
  }%
}
\makeatother

\AddToHook{begindocument/end}{%
}

\newcommand{\Ls}{L_{\mathrm{s}}}

\newcommand{\Dt}{\Delta t}
\newcommand{\Dphi}{\Delta\varphi}
\newcommand{\ndot}{\dot{n}}
\providecommand{\degr}{\ensuremath{^\circ}}
\newcommand{\kms}{\ensuremath{\,\mathrm{km\,s^{-1}}}}
\newcommand{\degs}{\ensuremath{^\circ}}

\newcommand{\asun}{\theta_\odot}
\newcommand{\aph}{\theta_{\mathrm{Ph}}}
\newcommand{\ade}{\theta_{\mathrm{De}}}

\begin{document}

\title{Simultaneous Double Transits of Phobos and Deimos \\
  as Seen from the Martian Surface: \\
  A Millennium Catalogue}

\author{Samuel Cody}
\email{sfjcody@gmail.com}

\begin{abstract}
I present the first systematic catalogue of simultaneous solar transits
of both Phobos and Deimos as observed from the surface of Mars.
Using the JPL mar099 ephemeris (Brozovi\'{c} et al.\ 2025) and SPICE
toolkit, I searched the millennium 1600--2600\,CE for epochs at which
both Martian moons project onto the solar disc at the same instant for
at least one surface location.  I identify 8565 near-miss double
transits, 49~partial-overlap double transits (both moons simultaneously
on the solar disc, with at least one partially cut off by the limb),
and 17~full double transits in which
both moons lie wholly within the solar disc at the same moment.  All
events cluster tightly around the Martian equinoxes
($\Ls \approx 0\degs$ or $180\degs$) and within $\pm 9\degs$ of the
equator, reflecting the near-equatorial orbital inclinations of both
moons.  I derive a hard theoretical latitude limit of $\pm 13.1\degs$
beyond which simultaneous double transits are geometrically impossible.
The next observable partial double transit, excluding less
prominent near-miss events, is predicted for 2034~April~17.
The next full double transit, with both moons wholly inside the solar
disc, with a gap between each silhouette and the solar limb, occurs
on 2118~November~20. The geometries for both these events were confirmed 
with JPL Horizons.
I provide uncertainty estimates based on the Brozovi\'{c} et al.\
covariance model, with predicted position errors growing from
$\sim$1\,km for near-term events to $\sim$600\,km at the catalogue
boundaries, and note that the JAXA MMX mission ($\sim$2031) will
dramatically reduce uncertainties for all post-2030 predictions.
\end{abstract}


\section{Introduction} \label{sec:intro}

The very earliest conception of what eventually became Frank Herbert's
seminal science fiction novel \textit{Dune} was set, not on Arrakis, a
world around a distant star, but on our own Mars.  This was quickly
abandoned, due to ``readers [having] too many preconceived ideas about
that planet, due to the number of stories that had been written about
it'' \citep{herbert2021}, but relics of this initial concept remain,
like Arrakis' tiny polar caps, and its dual moons.

These two moons are not given Imperial names in any of the six
\textit{Dune} novels, but in the \textit{Dune Encyclopedia}
\citep{mcnally1984}, semi-canonised by Frank Herbert before his death,
they are referred to as Krelln and Arvon, Arvon being the smaller, but closer
``second'' of the two, and hosting the famous kangaroo mouse-like
markings that give it its more widely known Fremen name, Muad'Dib.
It is said in the \textit{Dune Encyclopedia} that Arvon can occult
Krelln, and does so regularly.

We see an example of the two moons in close visual proximity in the
opening scene of Denis Villeneuve's \textit{Dune: Part Two} (2024).
The moon with the larger apparent diameter is seen eclipsing the 
Arrakian sun, canonically Canopus (Appendix~I of \citealt{herbert1965}),
while the other moon is transiting it.  The wan eclipse light makes for a
breathtaking spectacle.  Villeneuve took advantage of a real, terrestrial partial
eclipse (2022 October~25), which took place during the film shoot in
Jordan; visual effects supervisor Paul Lambert later composited a
second moon into the footage.

The question naturally arises: can anything even half as spectacular
take place on our own Mars, inspiration for Arrakis?  And if so,
where and when?

\bigskip

SPICE kernels are specialised data files developed by NASA's Navigation
and Ancillary Information Facility \citep[NAIF;][]{Acton2018} that
provide essential geometry and ancillary information for space missions.
They contain data on spacecraft trajectory, orientation, and planetary
ephemerides, enabling precise calculation of instrument pointing and
observation geometry.  The most recent such kernel for Mars' two moons,
Phobos and Deimos, is \texttt{mar099.bsp}.  This is based on the
revised Martian satellite ephemerides of \citet{Brozovic2025}, which
incorporate over a century of astrometric observations, from
historical ground-based measurements through modern spacecraft imaging
and eclipse-timing data from Mars Express, MRO, and the Perseverance
rover's Mastcam-Z.  It supersedes the widely used \texttt{mar097}
ephemeris that had been the standard for over a decade, and covers
1600--2600\,CE with substantially improved constraints on the tidal
acceleration of Phobos.

Solar transits of the Martian moons, in which Phobos or Deimos passes
across the solar disc as seen from the planet's surface, have been
observed by every NASA surface mission since Spirit and Opportunity
\citep{Bell2005}. Even as far back as 1977, Viking 1 was able to detect
and measure the slight dimming resulting from a Phobos transit,
though imagery of the transit itself was not produced.
These events serve as astrometric calibration
points for lunar ephemerides \citep{Jacobson2014, Brozovic2025} and
provide constraints on the tidal evolution of the Mars system
\citep{Lainey2007, Lainey2021}.

Phobos transits are dramatic: the moon subtends roughly $0.1\degs$ in
diameter (about one-third the angular diameter of the Sun from Mars)
and races across the disc in 20--30\,s.  Deimos transits are more
subtle, the moon appearing as a small dark speck roughly $0.034\degs$
across (one-tenth of the solar diameter), drifting slowly westward over
1--2\,min.  Both phenomena have been imaged extensively by the
Pancam, Mastcam, and Mastcam-Z instruments
\citep{Bell2005, Lemmon2013, Bell2017}.

Whether both moons can ever transit the Sun \emph{simultaneously} as
seen from a single surface location appears to be unaddressed in the
literature.  Casual inspection suggests this
should be rare: Phobos and Deimos have different orbital periods
(7.66\,h and 30.3\,h), different orbital inclinations ($1.08\degs$ and
$2.68\degs$ to the Laplace plane), and their shadow tracks move in
opposite directions across the surface.  Whether the geometry ever
aligns, and if so, how often and for what class of observer, has not
previously been computed.

In this paper I carry out that computation using the mar099 ephemeris
introduced above.  I search the full
millennium 1600--2600\,CE and classify
every simultaneous double transit by the depth of overlap.  I derive
the theoretical latitude limits, quantify the seasonal clustering, and
provide a catalogue with uncertainty estimates suitable for planning
future observations, whether by robotic landers, rovers, or eventual
human explorers.

\section{Geometry and Definitions} \label{sec:geometry}

\subsection{Single Transit Geometry} \label{sec:single}

A transit occurs when a moon's areocentric angular separation from the
Sun centre falls below the sum of their angular radii:
\begin{equation}
  \rho < \asun + \theta_{\mathrm{moon}},
  \label{eq:transit_threshold}
\end{equation}
where $\asun \approx 0.175\degs$ is the angular radius of the Sun as
seen from Mars, $\aph \approx 0.050\degs$ for Phobos, and
$\ade \approx 0.017\degs$ for Deimos.  For a \emph{full} transit
(moon entirely within the disc), the condition is
$\rho < \asun - \theta_{\mathrm{moon}}$.

The shadow of each moon sweeps a track across the Martian surface.
Phobos, orbiting faster than Mars rotates, casts a shadow that moves
\emph{eastward} at $\sim$2\kms; Deimos, orbiting slower than Mars
rotates, produces a shadow drifting \emph{westward} at $\sim$0.5\kms.
The partial-transit track width is $\sim$59\,km for Phobos and
$\sim$134\,km for Deimos; the full-transit track widths are
$\sim$15\,km and $\sim$110\,km respectively.  A surface observer's
latitude determines which shadow tracks pass overhead.

\subsection{Observer Parallax} \label{sec:parallax}

From the areocentre, both moons always transit at the same apparent
latitude as their shadow centre.  A surface observer, however, is
displaced from the centre of mass: for an observer at areographic
latitude $\varphi$, the parallax displacement shifts the apparent
position of each moon by an amount that depends on the ratio of Mars's
radius to the moon's orbital distance.  The parallax lever for Phobos
($a = 9375$\,km) is
\begin{equation}
  \frac{R_{\mathrm{Mars}}}{a_{\mathrm{Ph}} - R_{\mathrm{Mars}}}
    \approx 0.568 \; \degs/\degs,
\end{equation}
and for Deimos ($a = 23{,}458$\,km) it is
\begin{equation}
  \frac{R_{\mathrm{Mars}}}{a_{\mathrm{De}} - R_{\mathrm{Mars}}}
    \approx 0.169 \; \degs/\degs.
\end{equation}
The \emph{differential} parallax between the two moons is therefore
$\sim$0.399\degs\ per degree of observer latitude.  This is the
fundamental geometric constraint: moving the observer north or south
shifts the two moons' apparent positions by different amounts, allowing
shadow tracks that are separated at the areocentre to overlap for a
suitably positioned surface observer; or conversely, causing tracks
that overlap at the areocentre to separate.

\subsection{Theoretical Latitude Limit} \label{sec:latlimit}

I derive the maximum observer latitude at which a simultaneous double
transit can occur.  The key parameters are:

\begin{enumerate}
  \item Phobos orbital inclination: $i_{\mathrm{Ph}} = 1.08\degs$
    $\Rightarrow$ orbital latitude range $\pm 1.09\degs$
  \item Deimos orbital inclination: $i_{\mathrm{De}} = 2.68\degs$
    $\Rightarrow$ orbital latitude range $\pm 2.68\degs$
  \item Sub-solar declination range: $\pm 25.2\degs$ (Mars obliquity)
  \item Differential parallax: $0.399\;\degs/\degs$
\end{enumerate}

In the optimal configuration---Phobos at maximum orbital latitude
($+1.09\degs$), Deimos at minimum ($-2.68\degs$), and the sub-solar
point at $\delta_\odot = -5.49\degs$ (so that the Deimos shadow track
is shifted poleward while Phobos's remains near the equator)---the
two shadow centres are separated by their maximum useful amount.  An
observer at latitude $\varphi$ can then use differential parallax to
bring both moons onto the disc simultaneously, provided the resulting
apparent separations both satisfy Equation~\ref{eq:transit_threshold}.

Solving this system yields a hard limit of
\begin{equation}
  |\varphi_{\mathrm{max}}| = 13.1\degs \quad (\pm 774\;\mathrm{km}).
  \label{eq:lat_limit}
\end{equation}
At this extreme, the overlap band narrows to $\sim$0.25\degs\ in
latitude (15\,km), making observation effectively impossible in
practice.  The observed catalogue range of $-8.7\degs$ to $+8.5\degs$
represents 66\% of the theoretical maximum, consistent with the fact
that the optimal orbital geometry rarely coincides with the required
solar declination.

Any future Mars colony or outpost situated above $\pm 13\degs$
latitude would never witness a simultaneous double transit, regardless
of how long its inhabitants waited.

\section{Search Methodology} \label{sec:method}

\subsection{Ephemeris} \label{sec:ephemeris}

All computations use the JPL mar099.bsp satellite ephemeris
\citep{Brozovic2025}, which provides Chebyshev polynomial
representations of Phobos and Deimos states over 1600--2600\,CE.  This
ephemeris is fitted to spacecraft tracking, imaging, and eclipse-timing
data accumulated through 2019, and includes a tidal acceleration
$\ndot/2 = (1.258 \pm 0.058) \times 10^{-3}\;\mathrm{deg\,yr^{-2}}$
for Phobos.  Planetary positions are from de440.bsp
\citep{Park2021}; Mars orientation from pck00011.tpc.
Computations use the SPICE toolkit via the \texttt{spiceypy} Python
binding \citep{Annex2020}.

\subsection{Search Algorithm} \label{sec:algorithm}

The search proceeds in three phases:

\paragraph{Phase~1: Conjunction Timing.}
For each Deimos superior conjunction (areocentric angular separation
from the Sun $< 0.192\degs$), I identify all Phobos conjunctions
occurring within $\pm$30\,min.  Conjunction times are found by
evaluating the areocentric moon--Sun angular separation at 2-second
intervals and refining minima to 0.1-second precision.  Over the full
search interval, this yields $\sim$10$^5$ Phobos--Deimos conjunction
pairs.

\paragraph{Phase~2: Shadow Geometry.}
At each conjunction time, SPICE computes the exact shadow-centre
latitude and longitude on the Martian surface by ray-tracing from the
moon through the parallel solar illumination to the reference
ellipsoid.  The shadow track width (partial and full) is computed from
the instantaneous Moon altitude and angular radii.

\paragraph{Phase~3: Overlap Classification.}
For each conjunction pair, I compute the latitude separation
$\Dphi$ between the Phobos and Deimos shadow centres and compare it to
the sum of track half-widths.  Events are classified as:

\begin{description}
  \item[Edge-only (E):] The partial-transit tracks overlap only at
    their extreme edges; an observer in the narrow overlap zone sees
    one moon well on the disc while the other narrowly misses the solar
    disc.  In practice the near-miss moon is barely discernible against
    limb brightening, and the event is visually indistinguishable from
    a single transit.  These marginal geometries account for the vast
    majority of double transits (8565 of 8631 events).
  \item[Partial overlap (P):] Both moons are simultaneously and
    unambiguously superimposed on the solar disc for at least one
    observer location, but the geometry requires a positional
    compromise; at least one moon is only partially on the disc.
    Unlike the marginal E class, a P double transit is a
    recognisable two-body event: both silhouettes are plainly visible
    against the photosphere, though one or both may be cut off by the
    limb.
  \item[Full overlap (F):] Both moons lie fully within the solar
    disc simultaneously.  Two dark silhouettes are visible against the
    photosphere.  In the rarest and most visually striking
    configuration, both silhouettes are comfortably framed with
    clear photosphere on all sides.  This requires
    $\Dt < 25$\,s (less than one Phobos transit duration), full
    shadow-track overlap, and latitude separation
    $\Dphi < 1.2\degs$.
\end{description}

\noindent
The conjunction time gap $\Dt$ between the two moons is the primary
discriminant: a Phobos transit lasts $\sim$25\,s from any fixed point,
so $\Dt \ll 25$\,s ensures the two transits are genuinely
simultaneous rather than sequential.  Sequential double transits, in
which one moon finishes crossing the disc before the other begins, are
far more common but are not considered here; the present catalogue
includes only events in which both silhouettes are on the solar disc
at the same instant.

\subsection{Validation with JPL Horizons} \label{sec:horizons}

Near-term predictions were independently verified using the JPL
Horizons on-line ephemeris service
\citep{Giorgini1996}\footnote{\url{https://ssd.jpl.nasa.gov/horizons/}}.
For the 2034~April~17 event, a grid search over surface observer
positions confirmed that a location at $(\varphi, \lambda) = (-0.92\degs,
160.9\degs\,\mathrm{E})$ observes Phobos at $\rho_{\mathrm{Ph}} =
0.123\degs$ and Deimos at $\rho_{\mathrm{De}} = 0.157\degs$ from the Sun
centre, both well within the transit threshold, at the same UTC epoch.
Since Horizons reads the same mar099 ephemeris, this constitutes a
verification of my parallax and ray-tracing implementation rather than
an independent dynamical confirmation.

\section{Results} \label{sec:results}

\subsection{Census} \label{sec:census}

Table~\ref{tab:summary} summarises the full 1600--2600\,CE search.
Of the $\sim$10$^5$ Phobos--Deimos conjunction pairs within
$\pm$30\,min, the overwhelming majority have shadow tracks separated
by more than their combined widths.

\begin{deluxetable}{lrc}
\tablecaption{Double Transit Census, 1600--2600\,CE \label{tab:summary}}
\tablehead{
  \colhead{Class} & \colhead{Count} & \colhead{Description}
}
\startdata
E (edge only)        & 8565 & One moon narrowly misses disc      \\
P (partial overlap)  &   49 & Both on disc, $\geq$1 partial \\
F (full overlap)     &   17 & Both fully on disc         \\
\hline
Total                & 8631 &                            \\
\enddata
\tablecomments{The 17~F events include one---2118 November~20---in
  which both silhouettes are comfortably framed within the disc.}
\end{deluxetable}

\subsection{Seasonal Clustering} \label{sec:seasonal}

\begin{figure}
\centering
\includegraphics[width=\columnwidth]{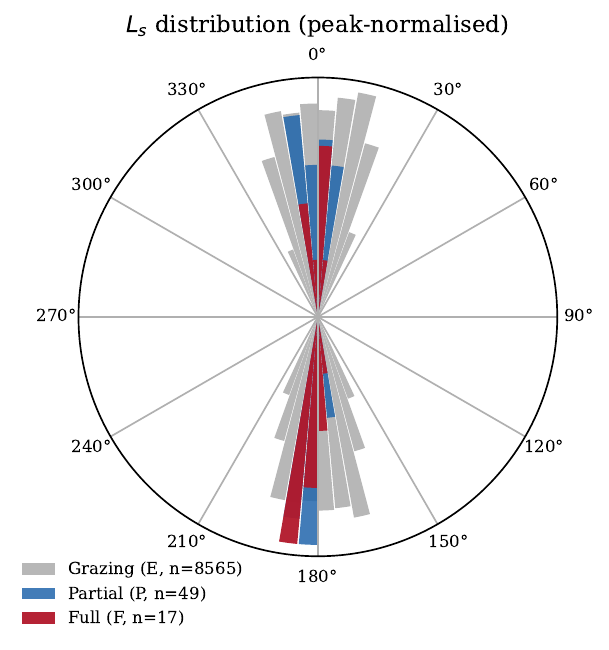}
\caption{Distribution of simultaneous double transits in areocentric
  solar longitude $L_s$.  All three overlap classes---near-miss (E),
  partial (P), and full (F)---cluster tightly around the Martian
  equinoxes ($L_s \approx 0\degs$ and $180\degs$), reflecting the
  near-equatorial orbital planes of both moons.  Each histogram is
  peak-normalised.
  \label{fig:ls}}
\end{figure}

All 17 full-overlap events occur within $\pm 9\degs$ of a Martian
equinox ($\Ls \approx 0\degs$ or $\Ls \approx 180\degs$;
Figure~\ref{fig:ls}).  This is a necessary consequence of the geometry:
at equinox, the sub-solar point crosses the equator where both moon
orbits are concentrated (inclinations $1.08\degs$ and $2.68\degs$),
maximising the probability that both shadow tracks lie at similar
latitudes.  Away from equinox, the sub-solar declination shifts both
shadow tracks poleward, but by different amounts due to the different
orbital inclinations, increasing the latitude gap beyond the threshold
for simultaneous transit.

The E events show a broader but still equinox-dominated distribution,
with a secondary concentration at solstice for events involving the
most extreme combinations of orbital latitude.

\subsection{Latitude Distribution} \label{sec:latdist}

The 17 full-overlap (F) events span areographic latitudes from
$-8.8\degs$ to $+8.5\degs$.  The 49 P events extend to
$\pm 10.4\degs$.  No events of any class exceed the theoretical limit
of $\pm 13.1\degs$ derived in Section~\ref{sec:latlimit}.  The
distribution is roughly symmetric about the equator, consistent with
the near-zero mean inclination of both moon orbits relative to the
Laplace plane.

\subsection{Full-Overlap Catalogue} \label{sec:catalog}

I identify 17 events in which both Phobos and Deimos lie fully within
the solar disc simultaneously; these are listed in full in Appendix
Table~\ref{tab:fc_fo_full}.  An additional 49 partial-overlap (P)
events are catalogued in Appendix Table~\ref{tab:po_full}.
Figure~\ref{fig:map} shows the geographic distribution of all F
and P events on the Martian surface, with $2\sigma$ position
uncertainty ellipses.  The figure displays the 1800--2200\,CE subset
for clarity, as events near the 2019 data arc have the lowest
positional uncertainties.

\begin{figure*}[p]
\centering
\thispagestyle{empty}%
\rotatebox{90}{%
  \begin{minipage}{1.03\textheight}
    \centering
    \vspace*{-0.8cm}%
    \includegraphics[width=\linewidth]{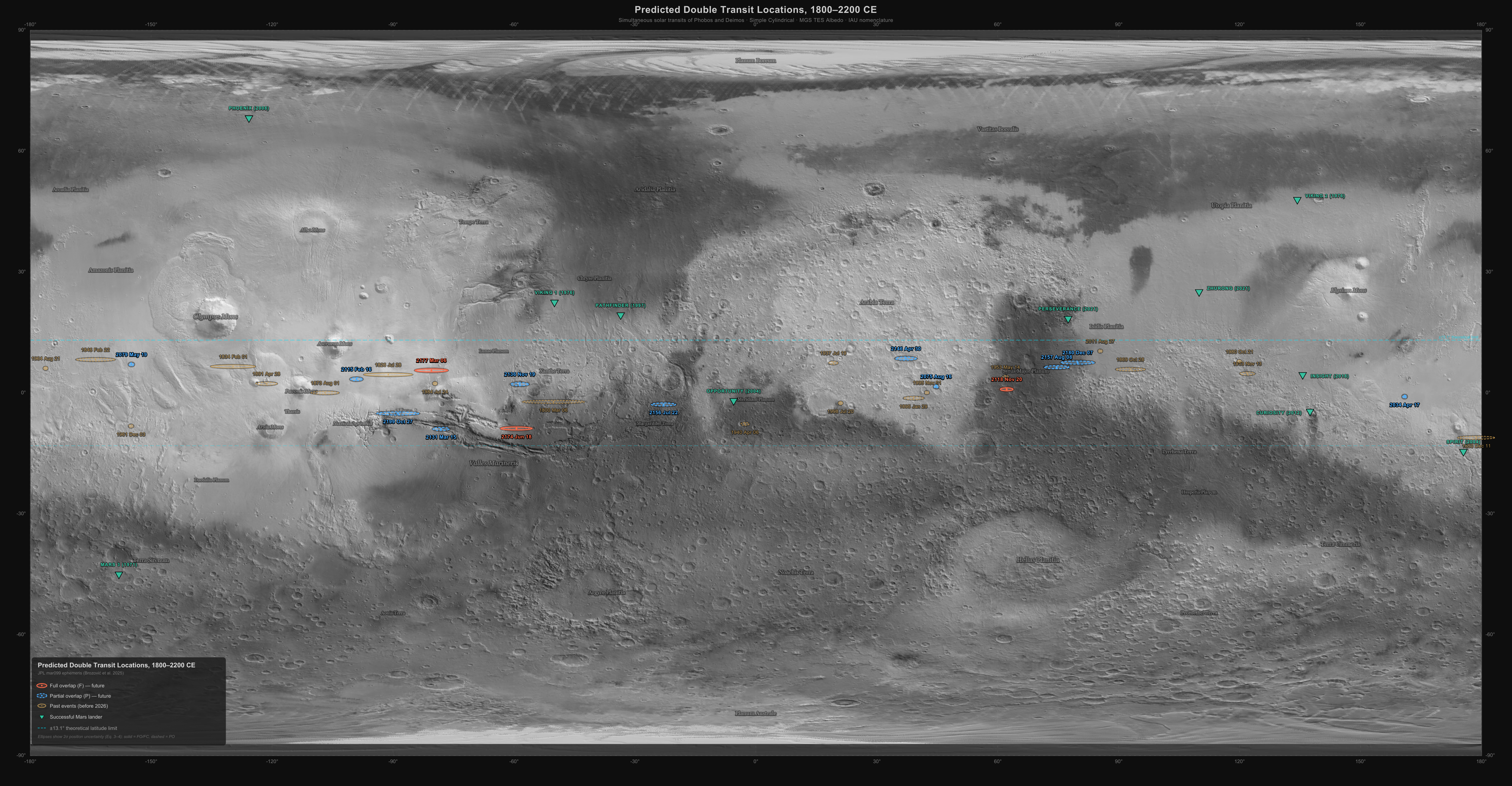}
    \caption{Geographic distribution of predicted simultaneous double
      transits of Phobos and Deimos on the Martian surface, 1800--2200\,CE,
      shown on an MGS TES Lambert Albedo basemap in Simple Cylindrical
      projection.  Ellipses indicate $2\sigma$ position uncertainties
      propagated from the Brozovi\'{c} et al.\ (2025) covariance model
      (Equations~\ref{eq:sigma_lon}--\ref{eq:sigma_lat}):
      events near the 2019 data arc have sub-km errors (nearly circular
      markers), while those at the catalogue boundaries show strongly
      elongated east--west ellipses reflecting the dominant along-track
      (longitude) uncertainty.  Solid ellipses denote full-overlap
      (F) events; dashed ellipses denote partial-overlap (P).
      Coral-red and blue mark future events; amber marks past events.
      Green triangles show the locations of all successful Mars landers.
      The dashed cyan band marks the $\pm 13.1\degs$ theoretical
      latitude limit.
      See Appendix Tables~\ref{tab:fc_fo_full} and~\ref{tab:po_full}
      for the full catalogues.
      \label{fig:map}\recordmappage}
    \vspace*{-0.3cm}%
    \hfill\pageref{fig:map}
  \end{minipage}%
}%
\end{figure*}

\subsection{Notable Events} \label{sec:notable}

\paragraph{2034 April 17 (P, $\Dt = 58$\,s).}
The next double transit of any kind.  This is a partial overlap:
an observer at $(-0.92\degs, 160.9\degs\,\mathrm{E})$ sees Phobos well
within the disc while Deimos clips the limb.  The event falls within
the high-precision regime of the mar099 ephemeris
($\sigma < 2$\,km) and has been validated via Horizons grid search.
No current or planned surface mission is positioned to observe it.

\paragraph{2118 November 20 (F, $\Dt = 12$\,s).}
The only full double in the next hundred years.  With a conjunction
gap of just 12\,s, less than half a Phobos transit duration, the
two shadows arrive nearly together.  The observer at
$(+0.90\degs, 62.1\degs\,\mathrm{E})$ would see both dark silhouettes
floating against the photosphere simultaneously, each well clear of
the solar limb.  Phobos appears as a large potato-shaped blot roughly
one-third the Sun's diameter; Deimos as a much smaller speck about
one-tenth.  This is the ``postcard'' double transit, the most visually
striking configuration that the Mars system can produce.

The location corresponds to the southwestern edge of Syrtis Major
Planum, a low-albedo volcanic plateau.  The position uncertainty at
$t = 99$\,yr from the 2019 data arc end is $\sigma_\lambda \approx
48$\,km, $\sigma_\varphi \approx 2$\,km.  The event itself is certain;
the only question is the exact standing point.

\section{Uncertainty Analysis} \label{sec:uncertainty}

\subsection{Error Budget} \label{sec:errbudget}

The dominant uncertainty in predicted transit locations is the
along-track (longitude) error, driven by the imperfectly known tidal
acceleration of Phobos.  Following \citet{Brozovic2025}, I adopt
their covariance model with realistic scaling factors ($4\times$ for
Phobos, $2\times$ for Deimos) and propagate uncertainties as:

\begin{equation}
  \sigma_\lambda(t) = \sqrt{
    \left(\frac{1.4\,t}{21}\right)^2 +
    \left(a_{\mathrm{Ph}} \cdot \tfrac{1}{2}\,\sigma_{\ndot} \cdot
    t^2\right)^2} \;\mathrm{km},
  \label{eq:sigma_lon}
\end{equation}
\begin{equation}
  \sigma_\varphi(t) = \sqrt{(0.01\,t)^2 + (0.015\,t)^2}\;\mathrm{km},
  \label{eq:sigma_lat}
\end{equation}
where $t$ is years since the end of the 2019 data arc.  The quadratic
term in Equation~\ref{eq:sigma_lon} encodes the tidal acceleration
uncertainty $\sigma_{\ndot/2} = 0.058 \times 10^{-3}$\,deg\,yr$^{-2}$.

Table~\ref{tab:uncertainty} gives representative values.  The
cross-track (latitude) error remains small even at long extrapolations,
because it depends on orbital inclination knowledge rather than mean
motion.  This means the \emph{existence} of predicted events is robust
even when the along-track error is large: an event with
$\sigma_\lambda = 600$\,km has uncertain longitude but a well-determined
latitude, and since the shadow track spans $360\degs$ of longitude during
each orbital period, the transit will occur \emph{somewhere} on Mars.

\begin{deluxetable}{lcccc}
\tablecaption{Representative Position Uncertainties (1$\sigma$)
  \label{tab:uncertainty}}
\tablehead{
  \colhead{Year} & \colhead{$t$ (yr)} &
  \colhead{$\sigma_\lambda$ (km)} &
  \colhead{$\sigma_\varphi$ (km)} &
  \colhead{Regime}
}
\startdata
1600 & 419 &   833   &   7.6 & Provisional    \\
1700 & 319 &   483   &   5.7 & Provisional    \\
1800 & 219 &   228   &   3.9 & Provisional    \\
1900 & 119 &    67.5 &   2.1 & Moderate       \\
1964 &  55 &    14.8 &   1.0 & Moderate       \\
1998 &  21 &     2.5 &   0.4 & High precision \\
2019 &   0 &  $<$0.5 &$<$0.2 & Data arc       \\
2034 &  15 &     1.5 &   0.3 & High precision \\
2079 &  60 &    17.8 &   1.1 & Moderate       \\
2118 &  99 &    47.0 &   1.8 & Moderate       \\
2200 & 181 &   156   &   3.3 & Provisional    \\
2382 & 363 &   626   &   6.5 & Provisional    \\
2600 & 581 & $>$1000 &   10  & Order-of-magnitude \\
\enddata
\tablecomments{Propagated from the Brozovi\'{c} et al.\ (2025)
  covariance model with $4\times$ Phobos, $2\times$ Deimos scaling.
  Uncertainties grow symmetrically from the 2019 data arc endpoint.}
\end{deluxetable}

\subsection{Impact of the MMX Mission} \label{sec:mmx}

The JAXA Martian Moons eXploration mission, scheduled for launch in
2026 with Mars arrival in 2027, will orbit Phobos for $\sim$3 years
and return a surface sample.  The proximity operations will yield
astrometric constraints orders of magnitude more precise than
Earth-based observations, collapsing the tidal acceleration uncertainty
and resetting the data arc to $\sim$2031.  For all catalogue events
after that date, the position uncertainties in
Table~\ref{tab:uncertainty} should be regarded as conservative upper
bounds that will be superseded once MMX tracking data are incorporated
into a future ephemeris update.

\section{Discussion} \label{sec:discussion}

\subsection{Rarity of Double Transits} \label{sec:rarity}

Single Phobos transits visible from a fixed equatorial site occur
roughly twice per (Martian) year near equinox.  Single Deimos transits
are comparably frequent, though less conspicuous.  Full double transits
(F), by contrast, average only 17 per millennium across the
\emph{entire planet}; roughly once per 60\,yr.  From any single
location, the wait is far longer.

The scarcity arises from the compounding of three independent
constraints: (1) the conjunction time gap $\Dt$ must be small compared
to the Phobos transit duration ($\sim$25\,s); (2) the shadow track
latitude gap $\Dphi$ must be less than the sum of the full-transit
half-widths ($\sim$1\degs--2\degs); and (3) the observer must be on
the dayside at the correct longitude.  Constraint~(1) depends on the
commensurability of the two orbital periods; constraint~(2) on the
orbital latitude phasing and sub-solar declination; constraint~(3) on
the Mars rotation phase.  Their joint probability is very small.

\subsection{Why Equinox?} \label{sec:equinox}

The tight $\Ls$ clustering (Figure~\ref{fig:ls}) is not coincidental
but geometrically required.  Both moons orbit within a few degrees of
the Martian equatorial plane.  At solstice, the sub-solar declination
reaches $\pm 25\degs$, shifting both shadow tracks far from the equator;
however, the displacement differs for each moon because of their
different orbital distances and hence different parallax levers.
The latitude gap $\Dphi$ is minimised when the sub-solar point lies
near the equator, i.e.\ at equinox, because then the differential
parallax contribution from solar declination vanishes to first order.

The $\Ls \approx 180\degs$ (autumn equinox) events are slightly
favoured over $\Ls \approx 0\degs$ (spring equinox) in my catalogue,
though the asymmetry is not statistically significant given the small
sample size.

\subsection{Observability} \label{sec:observability}

None of the catalogue events coincide with a known or planned surface
mission location.  The 2034 event at $(160.9\degs\,\mathrm{E},
-0.92\degs)$ falls in the Elysium Planitia lowlands, roughly
1000\,km southeast of InSight's location, and no current mission
will be operational at the required coordinates.

A future rover or lander could, in principle, be tasked to drive to a
predicted observation point, provided the position uncertainty is
within the mission's traverse capability.  For the 2034 event,
the $\sim$2\,km uncertainty is well within the demonstrated range
of Curiosity and Perseverance.  For the 2118 event, the
$\sim$50\,km uncertainty would require mission planning but is not
infeasible for a future mobile platform.

Human explorers on Mars would face an additional constraint: the
predicted locations are in equatorial regions with no topographic
shelter, making them unattractive for habitation.  However, a
dedicated expedition to observe the 2118 event---analogous to 19th
century eclipse expeditions on Earth---would constitute a remarkable
milestone in human exploration.

\section{Summary} \label{sec:summary}

I have presented the first systematic search for simultaneous double
solar transits of Phobos and Deimos as seen from the Martian surface.
The principal findings are:

\begin{enumerate}
  \item Over the millennium 1600--2600\,CE, I identify 8565 near-miss
    (E), 49 partial-overlap (P), and 17 full (F) simultaneous double
    transits.

  \item All events cluster near the Martian equinoxes
    ($\Ls \approx 0\degs$ or $180\degs$) and within $\pm 9\degs$ of
    the equator, reflecting the near-equatorial orbital planes of both
    moons.

  \item The theoretical maximum observer latitude for a double transit
    is $\pm 13.1\degs$, set by the differential parallax between
    Phobos and Deimos.

  \item The next full double transit, on 2118 November~20, has a
    conjunction gap of only 12\,s and would present two dark
    silhouettes simultaneously and comfortably framed within the
    solar disc.

  \item The next double transit of any kind (partial overlap) is
    predicted for 2034 April~17 at
    $(-0.92\degs, 160.9\degs\,\mathrm{E})$, confirmed via JPL Horizons.

  \item Predicted positions have $\sigma \lesssim 2$\,km for near-term
    events, growing to $\sim$50\,km at a century; the JAXA MMX mission
    will dramatically reduce all post-2030 uncertainties.
\end{enumerate}

The full catalogue and search tools are available at
\url{https://doi.org/10.5281/zenodo.18665057}.

\vspace{1em}

\textit{Software:}
\texttt{spiceypy} \citep{Annex2020},
SPICE/NAIF Toolkit \citep{Acton2018},
JPL Horizons \citep{Giorgini1996}.

\textit{Data:}
mar099.bsp \citep{Brozovic2025},
de440.bsp \citep{Park2021}.

\begin{acknowledgments}

This research made use of NASA's Navigation and Ancillary Information
Facility (NAIF) SPICE toolkit and the JPL Horizons ephemeris service.
I thank M.\ Brozovi\'{c}, R.A.\ Jacobson, and R.S.\ Park for making
the mar099 satellite ephemeris available.

\end{acknowledgments}

\facilities{JPL Horizons}
\software{spiceypy \citep{Annex2020},
  SPICE Toolkit \citep{Acton2018},
  Python, NumPy, Matplotlib}

\appendix

\section{Full-Overlap Event Catalogue} \label{app:fc_fo}

Table~\ref{tab:fc_fo_full} lists all 17 full-overlap (F) simultaneous double transits identified in the
1600--2600\,CE search.

\startlongtable
\begin{deluxetable*}{rccrcrrccrrc}
\tablecaption{Complete Catalogue of Full Double Transits
  (F), 1600--2600\,CE
  \label{tab:fc_fo_full}}
\tablewidth{0pt}
\tablehead{
  \colhead{\#} &
  \colhead{Date} &
  \colhead{UTC} &
  \colhead{$L_s$} &
  \colhead{$\Delta t$} &
  \colhead{$\varphi_\mathrm{obs}$} &
  \colhead{$\lambda_\mathrm{obs}$} &
  \colhead{$\Delta\varphi$} &
  \colhead{Type} &
  \colhead{$\sigma_\lambda$} &
  \colhead{$\sigma_\varphi$} &
  \colhead{Tier} \\
  \colhead{} &
  \colhead{} &
  \colhead{} &
  \colhead{(\degr)} &
  \colhead{(s)} &
  \colhead{(\degr)} &
  \colhead{(\degr E)} &
  \colhead{(\degr)} &
  \colhead{} &
  \colhead{(km)} &
  \colhead{(km)} &
  \colhead{}
}
\startdata
  1 & 1674-03-08 & 06:51 & 178.8 & 12 & $-2.89$ & $28.6$ & 0.82 & F & 565.5 & 6.2 & A \\
  2 & 1715-08-09 & 03:22 & 187.1 & 58 & $+4.46$ & $353.0$ & 0.80 & F & 439.2 & 5.5 & B \\
  3 & 1725-10-30 & 01:25 & 3.5 & 30 & $-4.29$ & $16.6$ & 1.01 & F & 410.8 & 5.3 & A \\
  4 & 1785-02-23 & 16:19 & 177.7 & 46 & $-3.90$ & $27.9$ & 0.43 & F & 260.4 & 4.2 & A \\
  5 & 1826-07-28 & 19:12 & 186.6 & 18 & $+4.54$ & $268.8$ & 0.49 & F & 177.3 & 3.5 & A \\
  6 & 1834-02-01 & 08:09 & 184.7 & 1 & $+6.56$ & $230.3$ & 0.43 & F & 162.9 & 3.3 & A \\
  7 & 1964-08-21 & 07:40 & 352.4 & 88 & $+6.11$ & $183.8$ & 0.89 & F & 14.8 & 1.0 & C \\
  8 & 1980-10-21 & 10:23 & 187.0 & 85 & $+7.78$ & $119.9$ & 0.66 & F & 7.7 & 0.7 & B \\
  9 & 1984-07-24 & 16:52 & 186.0 & 88 & $+2.33$ & $280.4$ & 0.74 & F & 6.3 & 0.6 & B \\
  10 & 1998-07-20 & 03:49 & 2.7 & 49 & $-2.53$ & $21.0$ & 0.50 & F & 2.5 & 0.4 & A \\
  11 & 2118-11-20 & 00:15 & 355.5 & 11 & $+0.90$ & $62.2$ & 0.80 & F & 47.0 & 1.8 & A \\
  12 & 2174-06-18 & 09:58 & 172.2 & 63 & $-8.82$ & $300.6$ & 0.26 & F & 114.5 & 2.8 & B \\
  13 & 2177-03-06 & 01:33 & 352.9 & 115 & $+5.58$ & $279.5$ & 0.72 & F & 119.0 & 2.8 & C \\
  14 & 2327-09-10 & 01:47 & 0.6 & 47 & $-0.07$ & $218.0$ & 0.34 & F & 450.8 & 5.6 & A \\
  15 & 2546-12-01 & 00:16 & 180.4 & 52 & $+0.87$ & $347.9$ & 0.23 & F & 1318.8 & 9.5 & A \\
  16 & 2577-11-13 & 19:36 & 5.4 & 55 & $-6.99$ & $256.2$ & 0.49 & F & 1478.5 & 10.1 & A \\
  17 & 2597-09-15 & 14:01 & 182.2 & 43 & $+0.21$ & $271.2$ & 0.08 & F & 1586.4 & 10.4 & A \\
\enddata
\tablecomments{
  Predicted surface locations where both Phobos and Deimos
  simultaneously transit the solar disc, from SPICE mar099.bsp
  (Brozovi\'{c} et al.\ 2025).
  $L_s$: areocentric solar longitude
  (Allison \& McEwen 2000).
  $\Delta t$: time gap between inferior conjunctions.
  $\varphi_\mathrm{obs}$, $\lambda_\mathrm{obs}$:
  observer areographic latitude and east longitude.
  $\Delta\varphi$: latitude gap between shadow-track centres.
  Type: F = both moons fully within the solar disc.
  $\sigma_\lambda$, $\sigma_\varphi$: 1$\sigma$ observer position
  uncertainty from mar099 covariance propagation
  (formal $\times$4 Phobos, $\times$2 Deimos; \S\ref{sec:errbudget}).
  Tier: A = near-certain (score $\geq$ 75),
  B = likely ($\geq$ 50), C = possible ($\geq$ 25),
  D = unlikely ($<$ 25).
  Uncertainties grow symmetrically from the 2019 data arc endpoint;
  the MMX mission ($\sim$2031) will substantially reduce
  $\sigma$ for all post-2030 predictions.
}
\end{deluxetable*}

\clearpage
\section{Partial-Overlap Event Catalogue} \label{app:po}

Table~\ref{tab:po_full} lists all 49 partial-overlap (P)
simultaneous double transits identified in the 1600--2600\,CE search.

\startlongtable
\begin{deluxetable*}{rccrcrrccrrc}
\tablecaption{Complete Catalogue of Partial Double Transits (P),
  1600--2600\,CE \label{tab:po_full}}
\tablewidth{0pt}
\tablehead{
  \colhead{\#} &
  \colhead{Date} &
  \colhead{UTC} &
  \colhead{$L_s$} &
  \colhead{$\Delta t$} &
  \colhead{$\varphi_\mathrm{obs}$} &
  \colhead{$\lambda_\mathrm{obs}$} &
  \colhead{$\Delta\varphi$} &
  \colhead{Type} &
  \colhead{$\sigma_\lambda$} &
  \colhead{$\sigma_\varphi$} &
  \colhead{Tier} \\
  \colhead{} &
  \colhead{} &
  \colhead{} &
  \colhead{(\degr)} &
  \colhead{(s)} &
  \colhead{(\degr)} &
  \colhead{(\degr E)} &
  \colhead{(\degr)} &
  \colhead{} &
  \colhead{(km)} &
  \colhead{(km)} &
  \colhead{}
}
\startdata
  1 & 1603-08-07 & 09:12 & 8.1 & 172 & $-7.53$ & $89.3$ & 0.58 & P & 822.0 & 7.5 & C \\
  2 & 1612-02-26 & 13:08 & 187.1 & 232 & $+8.31$ & $232.4$ & 0.32 & P & 786.8 & 7.3 & C \\
  3 & 1614-11-12 & 22:13 & 5.0 & 203 & $-6.06$ & $294.7$ & 0.43 & P & 779.1 & 7.3 & B \\
  4 & 1631-09-21 & 19:21 & 352.2 & 293 & $+7.90$ & $243.0$ & 1.40 & P & 715.1 & 7.0 & D \\
  5 & 1691-12-03 & 18:00 & 354.6 & 157 & $+2.11$ & $12.7$ & 1.33 & P & 511.2 & 5.9 & D \\
  6 & 1702-05-29 & 02:49 & 180.7 & 267 & $+1.12$ & $14.1$ & 1.54 & P & 477.5 & 5.7 & D \\
  7 & 1719-05-12 & 09:52 & 186.1 & 46 & $+3.60$ & $153.5$ & 1.29 & P & 427.7 & 5.4 & B \\
  8 & 1723-02-12 & 16:21 & 185.1 & 69 & $+6.24$ & $314.0$ & 1.29 & P & 416.4 & 5.3 & C \\
  9 & 1762-08-11 & 07:53 & 184.1 & 233 & $+0.58$ & $60.4$ & 0.73 & P & 314.0 & 4.6 & C \\
  10 & 1788-11-26 & 22:46 & 176.7 & 123 & $-0.44$ & $188.7$ & 1.45 & P & 253.8 & 4.2 & D \\
  11 & 1800-03-06 & 18:17 & 173.9 & 167 & $-2.21$ & $309.8$ & 1.07 & P & 228.1 & 3.9 & C \\
  12 & 1846-02-22 & 07:47 & 352.5 & 49 & $+8.27$ & $196.2$ & 1.33 & P & 142.5 & 3.1 & B \\
  13 & 1850-12-11 & 05:09 & 170.1 & 51 & $-11.11$ & $178.7$ & 1.62 & P & 136.0 & 3.0 & C \\
  14 & 1869-10-28 & 23:34 & 185.3 & 183 & $+5.84$ & $92.9$ & 0.31 & P & 107.3 & 2.7 & C \\
  15 & 1873-08-01 & 06:04 & 184.3 & 275 & $+0.03$ & $253.1$ & 0.47 & P & 101.7 & 2.6 & C \\
  16 & 1891-04-28 & 17:01 & 359.6 & 193 & $+2.30$ & $238.6$ & 0.55 & P & 78.2 & 2.3 & C \\
  17 & 1895-01-29 & 23:32 & 358.8 & 93 & $-1.29$ & $39.1$ & 1.52 & P & 73.5 & 2.2 & D \\
  18 & 1913-11-13 & 07:57 & 354.6 & 148 & $+4.79$ & $121.9$ & 0.68 & P & 53.8 & 1.9 & C \\
  19 & 1937-07-19 & 23:39 & 187.6 & 293 & $+7.44$ & $19.2$ & 1.24 & P & 32.4 & 1.5 & D \\
  20 & 1940-04-06 & 15:15 & 6.4 & 272 & $-7.70$ & $357.3$ & 0.09 & P & 30.1 & 1.4 & C \\
  21 & 1953-05-14 & 12:15 & 355.0 & 196 & $+3.96$ & $61.9$ & 1.25 & P & 21.1 & 1.2 & D \\
  22 & 1961-12-03 & 16:05 & 171.8 & 105 & $-8.23$ & $205.0$ & 1.48 & P & 16.4 & 1.0 & D \\
  23 & 1995-11-01 & 12:16 & 183.2 & 189 & $+0.07$ & $42.4$ & 0.88 & P & 3.2 & 0.4 & C \\
  24 & 2011-08-27 & 00:49 & 351.1 & 28 & $+10.38$ & $85.4$ & 1.17 & P & 0.6 & 0.1 & A \\
  25 & 2034-04-17 & 01:32 & 3.9 & 58 & $-0.92$ & $160.9$ & 1.27 & P & 1.5 & 0.3 & B \\
  26 & 2075-08-16 & 07:10 & 355.3 & 184 & $+1.58$ & $44.7$ & 1.09 & P & 15.3 & 1.0 & D \\
  27 & 2079-05-19 & 13:39 & 354.5 & 237 & $+7.07$ & $205.1$ & 0.36 & P & 17.6 & 1.1 & C \\
  28 & 2115-02-16 & 17:51 & 356.3 & 285 & $+3.41$ & $260.9$ & 0.22 & P & 44.2 & 1.7 & C \\
  29 & 2131-03-15 & 17:01 & 172.0 & 198 & $-8.93$ & $281.9$ & 1.53 & P & 60.0 & 2.0 & D \\
  30 & 2136-11-19 & 04:23 & 180.1 & 292 & $+2.16$ & $301.4$ & 1.06 & P & 65.4 & 2.1 & C \\
  31 & 2146-04-30 & 04:44 & 188.2 & 284 & $+8.54$ & $37.2$ & 0.07 & P & 77.0 & 2.3 & C \\
  32 & 2156-07-22 & 09:11 & 5.5 & 222 & $-2.86$ & $337.1$ & 1.35 & P & 89.6 & 2.5 & D \\
  33 & 2157-08-08 & 06:41 & 186.0 & 252 & $+6.39$ & $74.6$ & 1.37 & P & 90.9 & 2.5 & D \\
  34 & 2180-12-07 & 08:02 & 352.0 & 88 & $+7.53$ & $79.8$ & 1.37 & P & 123.5 & 2.9 & C \\
  35 & 2199-10-27 & 08:40 & 6.3 & 167 & $-5.10$ & $271.2$ & 0.13 & P & 154.3 & 3.2 & C \\
  36 & 2203-07-31 & 15:10 & 5.4 & 167 & $-1.31$ & $71.7$ & 0.41 & P & 161.2 & 3.3 & B \\
  37 & 2208-05-19 & 19:01 & 185.0 & 219 & $+0.53$ & $330.5$ & 0.56 & P & 170.0 & 3.4 & C \\
  38 & 2215-11-24 & 07:55 & 183.1 & 27 & $+3.38$ & $292.1$ & 1.63 & P & 182.8 & 3.5 & B \\
  39 & 2218-08-11 & 23:27 & 2.8 & 147 & $-5.28$ & $271.0$ & 1.02 & P & 188.5 & 3.6 & D \\
  40 & 2255-05-28 & 00:47 & 184.9 & 84 & $+1.59$ & $66.5$ & 1.15 & P & 264.9 & 4.3 & C \\
  41 & 2306-03-09 & 19:33 & 184.6 & 221 & $+3.94$ & $238.8$ & 0.12 & P & 391.5 & 5.2 & C \\
  42 & 2382-03-26 & 09:19 & 0.2 & 35 & $+1.50$ & $108.2$ & 1.31 & P & 626.0 & 6.5 & B \\
  43 & 2523-04-27 & 23:24 & 4.6 & 169 & $-4.83$ & $172.0$ & 1.08 & P & 1206.3 & 9.1 & D \\
  44 & 2527-01-29 & 05:47 & 3.7 & 203 & $+0.26$ & $333.1$ & 0.60 & P & 1225.5 & 9.2 & C \\
  45 & 2543-02-27 & 17:51 & 181.3 & 284 & $-1.97$ & $186.8$ & 0.16 & P & 1303.9 & 9.4 & C \\
  46 & 2558-12-24 & 12:42 & 350.4 & 148 & $+10.93$ & $147.1$ & 1.33 & P & 1379.6 & 9.7 & D \\
  47 & 2572-03-08 & 01:50 & 357.5 & 154 & $+3.27$ & $320.4$ & 1.26 & P & 1452.1 & 10.0 & D \\
  48 & 2581-08-17 & 02:00 & 4.6 & 265 & $-1.06$ & $57.2$ & 0.55 & P & 1499.8 & 10.1 & C \\
  49 & 2592-11-26 & 10:18 & 3.3 & 236 & $+0.99$ & $11.6$ & 1.46 & P & 1559.0 & 10.3 & D \\
\enddata
\tablecomments{
  49 partial-overlap events extracted from the full 8614-event catalogue.
  Column definitions as in Table~\ref{tab:fc_fo_full}.
  The complete catalogue including all 8565 edge-only (E) events is
  available as a machine-readable table at \url{https://doi.org/10.5281/zenodo.18665057}.
}
\end{deluxetable*}

\vspace{1em}
\noindent\textit{v2: Renamed ``grazing'' double transits to ``near-miss''
throughout, to better reflect that these are cases where an initial pass
of the algorithm (Phase~3) flagged close proximity of both moons to the
Sun, but further inspection (Phase~4) revealed no true simultaneous
double transit.  No results were changed.}


\end{document}